\documentstyle[aps,prb,epsf]{revtex}
\tighten
\begin{document}
%\draft

\title{ \large \bf Universal Fluctuations of the Random Lasing Threshold
in a Sample of a Finite Area \\ }
\author{\large V. M. Apalkov and M. E. Raikh\vspace{1mm}}
\address{\large Department of Physics, University of Utah, Salt Lake City,
Utah 84112 \vspace{3mm} }
\maketitle

\begin{abstract}
\large
We consider the random lasing from a weakly  scattering
medium and demonstrate that the distribution of the
threshold gain over the ensemble of statistically independent
finite-size samples
is {\em universal}.
Universality stems from the
facts that: (i) lasing threshold in a given sample is determined
by the highest-quality mode of all the random resonators present in
the sample, and (ii)
 the  areal {\em density} of the random resonators decays
sharply  with the quality factor of the  mode that they  trap.
We find analytically the shape
of the universal distribution function of the lasing threshold.
The shape of this function is governed by a single
dimensionless parameter, $\beta$.
This parameter increases as a power law with $\ln S$, where $S$
is the sample area (length, volume),
and decreases as a power law with disorder strength. The powers
depend on the microscopic mechanism of the light trapping.
As a result, the distribution of the thresholds narrows with
$S$ and broadens with the disorder strength.
\end{abstract}


\Large

\centerline{\bf I. Introduction}
\vspace{5mm}

During the last five years the phenomenon of {\em coherent}
random lasing has been observed in a wide variety of disordered
media. These include semiconductor powders\cite{cao99},
polymer films\cite{frolov99},
dye-infiltrated opals\cite{frolov99}, etc. The fact,
that beyond a certain
optical excitation threshold, the random media emits a coherent light
(this fact was demonstrated in the photon statistics
experiments\cite{statistics1,statistics2})
suggests that certain disorder configurations in the medium (random
resonators\cite{we}) are
capable of ``trapping'' the light, thus assuming the role of the
conventional Fabry-Perot resonators. These configurations are sparse,
since on average the light propagation in the media is diffusive,
i.e. the condition $kl \gg 1$ is met, where $k$ is the wave number
and $l$ is the mean free path. As it was demonstrated analytically,
\cite{we} even for a {\em given}
$kl$--value, the ability of the medium to form resonators depends
strongly on the character
of the disorder. For example, from the point of view of trapping the
light, smooth disorder is much more favorable than the point-like
scatterers\cite{we} (see also Ref. \onlinecite{weBurin}).
It is also clear on the general grounds, that two
{\em statistically identical} disordered samples of a finite size
have different thresholds for coherent lasing. This is because
the most ``capable'' random resonators, that are present in each
sample and  determine its lasing threshold, have different
quality factors\cite{we1}. Therefore,  the study of the threshold
dependence on the sample size requires a statistical description.

Consider for concreteness a polymer film. Then the role of the
sample size is simply played by the area, $S$, of the  excitation
spot. Quantitatively, the change of the threshold excitation
intensity, $I$, with $S$ should be characterized by the evolution
of the distribution function, $F_S(I)$,
of the thresholds over many non-overlapping spots.
Namely, the position of maximum of $F_S(I)$, that describes the
threshold of a {\em typical} sample, should shift towards lower $I$
with increasing $S$,
whereas the width, that carries the information about the spread of
the thresholds in different samples, should decrease with $S$.

The main message of the
present paper is that the shape of the function $F_S(I)$ is
{\em universal} and is given by
\begin{equation}
\label{main}
F_S(I)= \frac{\beta_s}{I}\left(\frac{I}{I_S}\right)^{-\beta_s}
\exp\left[-\left(\frac{I}{I_S}\right)^{-\beta_s}\right],
\end{equation}
where the
%dependence of the
typical value, $I_S$,
is related to the sample area $S$ as follows
\begin{equation}
\label{typical}
I_S \propto \exp\left\{-\left[\frac{\ln\left(S/S_0\right)}{G}\right]
^{1/\lambda}\right\},
\end{equation}
where the parameters $\lambda > 1$ and $G$ are determined
by the {\em intrinsic}
properties of the disordered medium and are independent of $S$;
the area $S_0$ is a typical area of a random resonator. These two
parameters play a different role: while $\lambda$ is determined
exclusively by the {\em shape} of the disorder correlator, $G$
is a measure of the {\em disorder strength}.
The dependence of
$\beta_s$ on $S$ is  logarithmical, namely,
%$\beta_s \propto \ln^{1-\lambda^{-1}}\left(S/S_0\right)$.
$\beta_s \propto\Bigl[\ln\left(S/S_0\right)\Bigr]^
{\left(\lambda -1\right)/\lambda}$. Most importantly,
parameter $\beta_s$ {\em decreases} with the disorder
parameter $\left(kl\right)^{-1}$ as a power
law, i.e. for a weakly scattering medium we have $\beta_s \gg 1$.
The concrete value of the exponent
$\partial \ln \beta/\partial\ln(kl)$
depends on the microscopic properties of the disorder.

Experimentally, establishing the form of the distribution function
amounts
to  dividing of the measured values of  $I$ into ``bins'' and
constructing a hystogram from these bins.
In this regard, it is convenient to
fit the data to the theory using the {\em momenta} of the distribution,
calculated with the help of experimentally determined hystogram.
In particular, as it
easily follows from  Eq. (\ref{main}), the first moment is given by
\begin{equation}
\label{average}
\langle I \rangle  =
   I_S~\! \Gamma \!\left( 1-\beta_s^{-1} \right),
\end{equation}
where $\Gamma(x)$ is the $\Gamma$-function.
It can be also shown from Eq. (\ref{main}) that the average logarithm
of $I/\langle I \rangle$ is given by
\begin{equation}
\langle \ln I \rangle  - \ln \langle I \rangle
%\langle \ln \left(I/\langle I \rangle\right)\rangle
= \phi(\beta_s),
\label{phi0}
\end{equation}
where the function $\phi(u)$ is defined as
\begin{equation}
\label{phi}
 \phi(u)=
\frac{\gamma }{u} - \ln \left[\Gamma\left(1-u^{-1}\right)
\right].
\end{equation}
Here $\gamma \approx 0.5772$  is the Euler's constant.
The dependence $\phi(u)$ is shown in Fig. 1. It diverges as $\ln(u-1)$
in the limit $u\rightarrow 1$ and approaches zero  as
$\phi(u)\approx -\frac{\pi^2}{12 u^2}$
for $u\gg 1$.
The relations Eq. (\ref{average}) and Eq. (\ref{phi})
are sufficient to restore the parameters $I_S$ and $\beta_s$ from
experimentally measured hystogram.
% in the following way.
Indeed, calculating from
the data the values of $\langle I \rangle$ and $\langle \ln I \rangle$,
allows first to determine the parameter $\beta_s$ from Fig. 1, and
subsequently, the parameter $I_S$ from Eq. (\ref{average}).
Comparison of the experimental distribution with theoretical prediction
Eq. (\ref{main}) allows to test the basic underlying assumption about
the statistical independence of different random resonators. On the
quantitative level, it allows to determine $\lambda$, which is strongly
sensitive to the microscopic arrangement of the disorder-induced
resonator.

The distribution function Eq. (\ref{main}) is plotted in Fig. 2 for
three values of $\beta_s$. Firstly, we observe that for smallest
$\beta_s$ the distribution $F_S(I)$ is broad and strongly asymmetric.
It has a long tail towards the high thresholds and falls off abruptly
towards low thresholds. Our main qualitative prediction
is that in experiments on random lasers
 the distribution of thresholds over
samples must have a distinctive shape similar to that of
the curve 1 in Fig. 2.
Indeed, experimentally, the lasing is observed only when the disorder
is strong enough. The realistic values\cite{we} of the disorder
parameter $kl$ are $\sim 5$. On the other hand, strong disorder
translates into the values of the parameter $\beta_s$ that are $\sim 1$.

As disorder decreases, leading to the rise of the average threshold,
the distribution $F_S(I)$ becomes progressively narrow and symmetric.
For highest $\beta_s=6$ (curve 3 in Fig. 2) this distribution is
close to gaussian, $F_S(I)\propto
\exp\left[-\frac{\beta_s^2}{2}\left(\frac{I-I_S}{I_S}\right)^2\right]$.
We emphasize, that the distribution Eq. (\ref{main})
is derived under the assumption that all samples (excitation spots)
are {\em statistically independent}.
Therefore, experimental observation of a strongly
asymmetric  distribution of thresholds over samples would
be a confirmation that the underlying  disordered medium
is homogeneous {\em on average}, i.e.
it does not contain large-size fluctuations of technological origin
(larger than the sample size).
The remaining part of the paper is organized as follows.
In Sect. II we demonstrate that, with accuracy of a numerical factor,
the basic characteristics of the distribution
Eq. (\ref{main}) can be found from a simple reasoning. In Sect. III we
provide a rigorous derivation of Eq. (\ref{main}). In Sect. IV we
justify the validity of the basic assumption behind Eq. (\ref{main})
concerning the functional form of the density of random resonators.
In Sect. V. we discuss the tests of the applicability of the
distribution (\ref{main}) based on the relations between the different
cumulants. The details of the calculation of these cumulants are
presented in the Appendix.

\vspace{5mm}
\centerline{\bf II. Qualitative consideration}
\vspace{5mm}

The position of maximum and the width of the distribution function $F_S$
can be found  following the qualitative
consideration of Refs.\onlinecite{JETP,singapore,elsevier}.
Denote with $\rho (Q)$ the areal {\em density}
of disorder-induced resonators with the quality factor $Q$. This
density represents the statistical weight of disorder
configurations that are capable of trapping a mode with an anomalously
small relative linewidth, $Q^{-1} \ll 1$.
According to the definition, $\rho(Q)$ is the characteristics of
an {\em infinite} medium.
It describes the ability of the medium
to trap the light wave for anomalously long time
(namely, for $Q$ wave periods). Note, that $\rho(Q)$ is a meaningful
characteristics of the medium in two and three dimensions, where
the light propagation is diffusive on average.

 The function $\rho(Q)$ was calculated
analytically for certain models of disorder in
Refs. \onlinecite{we,review,weJosa}. The related quantity was also
invoked in Ref. \onlinecite{burin} for the analysis of
simulation results.
Similar function was studied in relation to current relaxation
in disordered conductors\cite{Kravtsov,Muzykantskii95,falko95,mirlin00}.
All analytical results obtained within different approaches and
for different models of disorder yield the same general form
of the density $\rho(Q)$ in the large-$Q$ limit, namely

\begin{equation}
\label{rho}
\rho (Q)\mbox{\Huge $|$}_{Q\gg 1} =
  S_0^{-1} \exp\left[-G\left(\ln Q\right)^{\lambda}\right],
\end{equation}
where, within a numerical factor, $G$ is related to the average
``conductance'',$kl$,  of the  disordered medium as $G = (kl)^{\mu}$,
with $\mu$ being a model-dependent exponent. The power, $\lambda$,
 of the $\ln Q$ in Eq. (\ref{rho}) does not depend on the $kl$-value.
It is determined exclusively by the form of the correlation function
of the disorder.

The position of the maximum of the threshold distribution follows from
a simple observation that the number of highest-$Q$ resonators,
responsible for lasing in a {\em typical} sample, is $\sim 1$. This
observation can be expressed analytically as
\begin{equation}
\label{observation}
S\rho(Q_S) \approx 1.
\end{equation}
Eq. (\ref{observation}) can be viewed as an equation for $Q_S$, which
is the quality factor of the highest-$Q$ resonator present in a typical
sample. Substituting the form (\ref{rho}) of $\rho (Q)$ into
Eq. (\ref{observation}), we obtain
\begin{equation}
\label{Qs}
Q_S = \exp \left\{ \left[\frac{\ln (S/S_0)}{G} \right]^{1/\lambda }
           \right\} .
\end{equation}

In order to estimate the width of the distribution, we introduce the
auxiliary quantity $\tilde{Q}_S$, defined as
 the maximal $Q$-factor of a resonator which is present in
almost all the samples. Since the density $\rho (Q)$ grows rapidly
with $Q$, the definition of $\tilde{Q}_S$ can be quantified by
the condition
\begin{equation}
\label{auxiliary}
S\rho(\tilde{Q}_S) \approx 2,
\end{equation}
meaning that if the number of resonators is $2$ {\em on average},
even with fluctuations $\pm 2^{1/2}$ taken into account such a resonator
will be present in a typical sample with a high probability.
From Eq. (\ref{auxiliary}) we readily find
\begin{equation}
\label{tilde}
\tilde{Q}_S = \exp \left\{ \left[\frac{\ln (S/2S_0)}{G} \right]^{1/\lambda }
           \right\} .
\end{equation}
Finally, it is reasonable to assume that maximal $Q$-factors
within the ensemble of samples lie within the interval between
$\tilde{Q}_S$ and $Q_S$,
which yields the following estimate for the distribution width in
the logarithmic scale
\begin{equation}
\label{logarithmic}
\ln Q_S - \ln\tilde{Q}_S \approx \frac{\ln 2 }{\beta _s} ,
\end{equation}
where
\begin{equation}
\beta_s = \lambda G^{1/\lambda }  \left[ \ln ( S/S_0)
     \right]^{1-\lambda^{-1}}  .
\label{beta}
\end{equation}

\vspace{8mm}
\centerline{\bf III. Derivation of Eq. (1)}
\vspace{5mm}

The disorder configurations constituting high-$Q$ resonators are
sparse, so that different resonators are statistically independent.
Then the distribution function of the number, $N(Q,S)$,
of resonators with
quality factor in the interval $\left[Q,Q+dQ\right]$  within
the area $S$ is Poissonian
\begin{equation}
\label{poisson}
P_{n_{\mbox{\tiny $Q$}}}(N) =
   \frac{n_{\mbox{\tiny $Q$}}^{\mbox{\tiny $N$}}
              e^{-n_{\mbox{\tiny $Q$}}}}{N!},
\end{equation}
where $n_{\mbox{\tiny $Q$}} = S\rho (Q)dQ$ is the average number
of resonators. Then probability, $F_S(Q)$, to find the resonator
with the maximum quality factor in the interval $\left[Q,Q+dQ\right]$
is determined by the expression
\begin{equation}
\label{F_1}
F_S(Q)dQ = P_{n_{\mbox{\tiny $Q$}}}(1) \prod_{Q_1>Q}
         P_{n_{\mbox{\tiny $Q_1$}}}(0),
\end{equation}
so that there is one resonator with quality factor from $Q$ to
$Q+dQ$ and there are no resonators with the higher quality
factor. Substituting Eq. (\ref{poisson}) into
Eq. (\ref{F_1}), we obtain
\begin{equation}
\label{F_2}
F_S(Q) dQ = n_{\mbox{\tiny $Q$}} e^{-n_{\mbox{\tiny $Q$}}}
 \prod_{Q_1>Q} e^{-n_{\mbox{\tiny $Q_1$}}} = n_{\mbox{\tiny $Q$}}
   \exp\left(  -\sum_{Q_1\geq Q}  n_{\mbox{\tiny $Q_1$}} \right).
\end{equation}
The summation in Eq.~(\ref{F_2}) goes over all the intervals
$\left[Q_1,Q_1+dQ\right]$ with $Q_1>Q$.
Replacing the sum in Eq.~(\ref{F_2}) by the integral, we get
the following expression for the distribution
function\cite{misirpashaev98}
\begin{equation}
\label{F_3}
F_S(Q) = S\rho(Q) \exp\left[-S\int_{Q}^{\infty}
      dQ_1 \rho(Q_1) \right] .
\end{equation}
It is easy to see that $F_S(Q)$ is normalized. To evaluate
the integral in Eq.~(\ref{F_3}), we take into
account that the density of resonators
is a rapidly decreasing function of $Q$. Then the main
contribution to the integral in Eq.~(\ref{F_3}) comes
from low limit, $Q_1 \approx Q$. With $\rho(Q)$ given by
Eq.~(\ref{rho}), the integral (\ref{F_3})
can be rewritten as
\begin{eqnarray}
& &  S\int_{Q}^{\infty}  dQ_1 \rho(Q_1) =
                            \frac{S}{S_0} \int_Q^{\infty } dQ_1
   \exp\left[-G \left(\ln Q_1 \right)^{\lambda }\right]
                                            \nonumber \\
   &  & = \frac{QS}{S_0} \int_{1}^{\infty } dq
   \exp\left[-G \left(\ln Q\right)^{\lambda }
                                  \left( 1+\lambda \frac{\ln q}{\ln Q}
   \right)\right]  = \left(\frac{QS}{\lambda G S_0} \right)
     \frac{\exp\left[ -G\ln^{\lambda }Q \right]}{\ln^{\lambda -1}Q } .
\label{integral}
\end{eqnarray}
%where the parameters $\beta_s $ and $Q_s$ are defined by
%Eqs.~(\ref{beta}) and (\ref{Qs}).
Substituting Eq.~(\ref{integral}) into Eq.~(\ref{F_3}),
%and taking into account Eqs.(\ref{beta})-(\ref{Qs}),
we obtain the following expression for $F_S(Q)$
\begin{equation}
\label{FQ1}
F_S(Q)=
\frac{S}{S_0} \exp\left\{
   - G \ln ^{\lambda }Q - \left(\frac{QS}{\lambda G S_0} \right)
     \frac{\exp\left[ -G\ln^{\lambda }Q \right]}{\ln^{\lambda -1}Q }
\right\}   .
\end{equation}
As it follows from the qualitative considerations
(Sect. II), the distribution function, $F_S(Q)$, has a sharp
maximum at $Q\approx Q_S \approx \exp \{ [ \ln (S/S_0)/G]^{1/\lambda }
\}$. For this reason we present $\ln Q$
in Eq.~(\ref{FQ1}) as $\ln Q = \ln Q_S + \ln (Q/Q_S)$,
where  $\ln (Q/Q_S)\ll \ln Q_S$. Then
Eq.~(\ref{FQ1}) takes the form
\begin{equation}
\label{FQ2}
F_S(Q)=\frac{S}{S_0} \left( \frac{Q}{Q_S}\right)^{-\beta_s} \!\!\!
 \exp\left\{
   - G \ln ^{\lambda }Q_S -  \left( \frac{Q}{Q_S}\right)^{-\beta_s}
   \left(\frac{SQ_S}{\lambda G S_0} \right)
     \frac{\exp\left[ -G\ln^{\lambda }Q_S \right]}{\ln^{\lambda -1}Q_S }
\right\}   ,
\end{equation}
where $\beta _s\gg 1$ is expressed through $Q_S$ as follows
\begin{equation}
\label{beta_1}
\beta _s = \lambda G \ln ^{\lambda -1} Q_S .
\end{equation}
It is seen from Eq.~(\ref{FQ2}) that $F_S(Q)$ assumes a simple form
\begin{equation}
\label{mainQ}
F_S(Q)=
\frac{\beta_s}{Q}\left(\frac{Q}{Q_S}\right)^{\beta_s}
\exp\left[-\left(\frac{Q}{Q_S}\right)^{\beta_s}\right]
\end{equation}
if $Q_S$ satisfies the relation
\begin{equation}
\label{Qs_1}
 \left(\frac{SQ_S}{\lambda G S_0} \right)
     \frac{\exp\left[ -G\ln^{\lambda }Q_S
   \right]}{\ln^{\lambda -1}Q_S } = 1 .
\end{equation}
On the other hand, this relation can be  rewritten as
\begin{equation}
\label{Qs_2}
G \ln ^{\lambda }Q_S - \ln \left(Q_S/\beta_s \right) = \ln (S/S_0).
%Q_S =
%\exp \left\{ \left[ \frac{ \ln (S/S_0) - \ln (\beta _s/Q_s)}{G}
%      \right]^{1/\lambda }  \right\}  .
\end{equation}
Since we have $G\gg 1 $, the second term in the l.h.s. can be
considered as a small correction. Neglecting this correction,
we immediately realize that $Q_S$, determined from the qualitative
consideration, indeed satisfies the relation (\ref{Qs_1}).
Upon substituting the value of $Q_S$, given by Eq.~(\ref{Qs}),
into Eq.~(\ref{beta_1}) we arrive to the final expression (\ref{beta})
for parameter $\beta_s$.

%Since $ \ln (\beta _s/Q_s) \ll \ln (S/S_0)  \sim G \ln^{\lambda }Q_S$
%the expression (\ref{Qs_2})  takes the form (\ref{Qs}).
%Then substituting this form for $Q_S$ into Eq.~(\ref{beta_1}),
%we obtain the expression (\ref{beta}) for the parameter
%$\beta_s$.

As the threshold gain, $I$, is proportional to $Q^{-1}$,
the distribution of thresholds over the samples
follows from Eq. (\ref{mainQ}) upon transformation
$F_S(Q)dQ= F_S(I)dI$ with $Q/Q_S = \left(I/I_S\right)^{-1}$.
Then $F_S(I)$ takes the form Eq. (\ref{main}).

Since the distribution $F_S(I)$ is broad, it is best
characterized by the moments of $\ln I$ defined as
\begin{equation}
\label{moments}
M_n =  \left\langle \ln^n \frac{I}{\left\langle I \right\rangle }
        \right\rangle   ,
\end{equation}
rather than by the moments of $I$. In Eq. (\ref{moments}) the
value $\left\langle I \right\rangle$ is the average threshold,
calculated from the distribution (\ref{main}). It is given by
Eq. (\ref{average}).
As it is shown in the Appendix, the moments $M_n$ can be calculated
analytically. They are  expressed through the derivatives of the
digamma function
 $\Psi^{(k)}  = \left. d^k\Psi(x)/dx^k \right|_{x =1} $.
The first three moments are given by the following expressions
\begin{equation}
M_1 = -\frac{\Psi(1)}{\beta_s} - \ln \left[ \Gamma
     \left(1-\beta_s^{-1} \right)\right]  ,
\label{M1}
\end{equation}
\begin{equation}
M_2 = \frac{\Psi^{(1)}}{\beta^2_s} + M_1^2 ,
\label{M2}
\end{equation}
\begin{equation}
M_3 = -\frac{\Psi^{(2)}}{\beta^3_s} + 3 M_1 M_2 -2 M_1^3 .
\label{M3}
\end{equation}
Note that Eq.~(\ref{M1}) yields Eqs.~(\ref{phi0})-(\ref{phi})
upon substituting $\Psi(1) = -\gamma $.

\vspace{8mm}
\centerline{\bf IV. Discussion}
\vspace{5mm}

In our consideration we have assumed that the areal density
of resonators with anomalously large
$Q$-values of the eigenmodes (sometimes in the
literature they are called quasimodes\cite{genack} or
quasistates\cite{statistics1}) has the form
$\rho(Q) \propto \exp\left[-G\left(\ln Q\right)^{\lambda}\right]$.
Below we argue that this form is {\em generic}. More precisely, the
arguments leading to Eq. (\ref{rho}) are applicable to all the
models of the two-dimensional
disorder\cite{we,mirlin00,vanneste,sebbah} considered in the literature.
These arguments are the following\cite{review,crossover}

\noindent (i) Denote with $A$ the characteristic area occupied
by the disorder configuration that traps the light during the
time $\omega^{-1}Q$. The fact that the resonator mode is
confined, while its frequency, $\omega$, is degenerate with the
continuum inevitably results in the evanescent losses and corresponding
limitation on the quality factor
\begin{equation}
\label{one}
\ln Q =\frac{\omega A^{1/2}}{c f\{\delta\epsilon(\mbox{r})\}},
\end{equation}
where we have roughly estimated the curvature of the perimeter
of the fluctuation as $A^{-1/2}$; the functional
$f\{\delta\epsilon(\mbox{r})\}$ describes the actual fluctuation
of the dielectric constant, $\delta\epsilon(\mbox{r})$, within the
resonator. It is important that regardless of the form of
$f\{\delta\epsilon(\mbox{r})\}$, it {\em decreases} with overall
magnitude of the fluctuation, $\overline{\delta\epsilon}$.

\noindent (ii) The phase volume of the fluctuation
trapping the light for more than $Q$ periods is zero.
Small statistical deviations of the dielectric constant from
$\delta\epsilon(\mbox{r})$ would  couple the trapped
mode to the propagating modes and cause its very fast leakage.
Therefore, in order to maintain the high quality factor, these
statistical deviations must be suppressed. The probability of
such a suppression can be estimated as follows. There are
${\cal M}=A\omega^2/4\pi^2c^2$ squares with a side equal to the
light wavelength within the area of the resonator. Each square
couples to the propagating modes {\em independently}. Thus
the resonator mode would leak out of a {\em typical} square
during the time $\tau_l = l/c$, where $l$ is the mean free path.
Hence, the probability that for each square this time exceeds
$\omega^{-1}Q$ is given by
\begin{equation}
\label{two}
{\cal P}=\left(\frac{Q}{\omega\tau_l}\right)^{-{\cal M}}
=\exp\left[-\frac{A\omega^2}{4\pi^2c^2}\ln\left(Q/kl\right)\right].
\end{equation}
Upon expressing $A$ from Eq. (\ref{one}) and substituting into
Eq. (\ref{two}) we obtain
\begin{equation}
\label{three}
{\cal P}=\exp\left[-f^2\{\delta\epsilon(\mbox{r})\}\left(\ln Q\right)^2
\ln\left(Q/kl\right)/4\pi^2\right]    .
\end{equation}

\noindent (iii) the rest of consideration is model dependent. This
is because in order to obtain the density of resonators that
``survive'' the statistical fluctuations, one has to average
Eq. (\ref{three}) over the model-dependent distributions
$\delta\epsilon(\mbox{r})$ of trapping configurations with some
weight $\exp\left[-{\cal F}\{\delta\epsilon(\mbox{r})\}\right]$.
Important is,
however, that $\cal P$ increases sharply with
$\overline{\delta\epsilon}$ due to the factor $f$, whereas the
probability of a trapping configuration,
$\exp\left[-{\cal F}\{\delta\epsilon(\mbox{r})\}\right]$,
 drops exponentially with increasing
$\overline{\delta\epsilon}$. Hence, we conclude that the density
of resonators
\begin{equation}
\label{four}
\rho(Q) = \int {\cal D}\{ \delta \epsilon \} ~\!
   \exp\left[-{\cal F}\{\delta\epsilon(\mbox{r})\}\right] ~\!
{\cal P} \{\delta\epsilon(\mbox{r})\}
\end{equation}
is determined by the saddle point of the functional integral
Eq. (\ref{four}). Finally, any reasonable forms of
$f$ and ${\cal F}$ (say, power-law) yields Eq. (\ref{rho}), since
in course of taking the saddle point, the power of $\ln Q$ in the
exponent reduces compared to Eq. (\ref{three}) and the large factor
$G(kl)$ emerges in front of the logarithm.

\vspace{8mm}
\centerline{\bf V. Conclusions}
\vspace{4mm}
\centerline{ {\large \bf A. General Remarks} }
\vspace{5mm}

(i) In the present paper we argue
that the dependence of a typical threshold on the sample size
is given by Eq. (\ref{typical}), and thus falls off with $S$
{\em slower} than the power law, in contrast to conclusion drawn
in Ref.\onlinecite{burin} on the basis of numerical simulations.
As follows from Eq. (\ref{typical}), the dependence of  $I_S$ on $S$
turns into a power law only in the limit $\lambda \rightarrow 1$.
In the most advanced analytical calculation\cite{review} for the
model of a smooth spatial fluctuations of the dielectric constant,
the value of parameter $\lambda$ was found to be $\lambda = 11/8$.
Existing experimental results\cite{cao01} on the dependence of the
threshold
on the sample area were fitted to the power-law.
If we neglect the difference between $\lambda^{-1}=8/11$ and $1$,
then the power-law
dependence emerging from Eq. (\ref{typical}) will assume the form
$I_S \propto \left(S/S_0\right)^{-1/G}$, where the disorder parameter
$G \gg 1$ is defined through Eq.~(\ref{rho}) and decreases with the
concentration of scatterers, $n$. Analytical consideration of
Ref. \onlinecite{review} yielded $G^{-1} \propto n^{3/4}$. The results
of numerical simulations of Ref. \onlinecite{burin} also suggest that
the exponent $G^{-1}$ increases with $n$, however it is impossible to
extract the actual dependence of $G^{-1}$ on $n$ from the numerics.

(ii) Consideration in the present paper was based on the assumption
that all random resonators, corresponding to a given disorder
configuration, have the same quality factor, $Q$, regardless of
their position within the sample. This is by no means the case,
if a {\em periodic modulation of the dielectric constant} is present
in the sample along with the disorder. In particular,
if the modulation is strong enough to create the photonic bandgap,
then $Q$-values for resonators located at the center of the
sample would be exponentially higher that for resonators close to
the perimeter. For this reason it is not surprising that in simulations
reported in Ref.~\onlinecite{yamilov04} the dependence of the threshold
on the sample size was found to be much faster than Eq. (\ref{typical}).
However, aside from expression (\ref{typical}) for $I_S$,
the general distribution (\ref{main}) should apply to the disordered
samples with underlying periodicity considered in
Ref.~\onlinecite{yamilov04}.

\vspace{8mm}
\centerline{ {\large \bf B. Universal Relation Between the Cumulants}}
\vspace{5mm}

On the qualitative level, experimental verification of the
distribution (\ref{main})
of the random lasing threshold, predicted in the present paper,
would be an observation of the characteristic asymmetry of the
measured hystogram (long tail towards high thresholds and abrupt
cutoff towards low thresholds). On the quantitative level, after
parameters $\langle I_S\rangle$ and $\beta_s$ are determined from
the hystogram, there still remains a question how accurately a
strongly asymmetric distribution
Eq. (\ref{main}) describes the experimental data.
There exists a regular procedure to test  quantitatively the agreement
between the theoretical and experimental distributions, when both are
broad. This procedure is based on the comparison of different
{\em cumulants} of $\ln I$ determined from the experimentally measured
hystogram. In particular, the second and the third cumulants must
satisfy a certain universal relation, which we derive below.

We define the cumulants of the distribution $F_S(I)$ as follows
\begin{equation}
\label{cumulants}
C_n = \left\langle \left[
\ln \frac{I}{\left\langle I \right\rangle } -
   \left\langle \ln \frac{I}{\left\langle I \right\rangle }
                                                   \right\rangle
\right]^n \right\rangle          ,
\end{equation}
so that they are directly related to the moments $M_n$
[Eq.~(\ref{moments})]. Then, upon rewriting $C_n$ in terms of $M_n$,
we obtain for the first two cumulants
\begin{equation}
C_2 = M_2 - M_1^2  = \frac{\Psi^{(1)}}{\beta^2_s} ,
\label{C2}
\end{equation}

\begin{equation}
C_3 = M_3 -2M_1M_2 + 2M_1^3 =  -\frac{\Psi^{(2)}}{\beta^3_s},
\label{C3}
\end{equation}
where the relations Eqs.~(\ref{M1})-(\ref{M3}) have been
used.  The higher order
cumulants can be found following the similar procedure
\begin{equation}
C_n = \left(-\frac{1}{\beta_s}\right)^n \Psi^{(n-1)}.
\label{Cn}
\end{equation}
Expressing disorder parameter $\beta _s$ through $C_2$ and
substituting the result into the expression (\ref{Cn}), we
obtain the universal relation between the cumulants of
the distribution (\ref{main})
\begin{equation}
\label{CnC2}
\frac{C_n}{C_2^{n/2}} = (-1)^n
                \frac{\Psi^{(n-1)}}{\left[\Psi^{(1)}\right]^{n/2}} .
\end{equation}
In particular, for $n=3$ Eq.~(\ref{CnC2}) yields the following
relation between
the third and the second cumulants
\begin{equation}
\label{relation}
\frac{C_3}{C_2^{3/2}} = -
                \frac{\Psi^{(2)}}{\left[\Psi^{(1)}\right]^{3/2}} =
   \frac{2^{5/2}3^{3/2}\zeta(3)}{\pi^3}  \approx 1.139,
\end{equation}
which are easiest to extract from the experimental data.

In conclusion of this subsection we note that
$C_2^{1/2}\approx  0.76/\beta_s$, which characterizes the width
of the distribution (\ref{main}), is close to the estimate
Eq.~(\ref{logarithmic}) obtained from qualitative consideration
in Sect.~II.

\vspace{8mm}
\centerline{ {\large \bf C.
 Relation to the Statistics of the Lyapunov Exponents
for the Fluctuation Tails in 1D}}
\vspace{5mm}

There are two major physical arguments that determine the shape
of the distribution Eq.~(\ref{main}). (i) The occurrence of a mode
with anomalously high $Q$ is a {\em rare} event; (ii) Different
random resonators are {\em statistically independent}.
Using these arguments, and the very general form Eq.~(\ref{rho})
of the average
density $\rho(Q)$ leads directly to the distribution
Eq.~(\ref{main}), as was demonstrated above.

Generality of the distribution Eq.~(\ref{main}) suggests that it
might be applicable to the different situations when the net
property of the finite-size sample is governed by the rare events.
One such situation is the {\em coherent} transmission of a disordered
1D chain of a given length, $L$. For electron energy in the ``body''
of the band the distribution of log-transmission, reduces to
the sum of random numbers, and, thus, is gaussian.
Deep in the tail of the density of states this distribution is
also close to gaussian, since the transmission, $T$, is dominated by
``under-barrier tunneling'' with a decrement weakly changing in space.
Interference processes do not play a role in this energy interval.
There exists, however, a range of energies close to the band-edge,
 where the central limit theorem is already not applicable, but
the interference processes are still relevant. The distribution of
the log-transmission becomes strongly asymmetric in this ``fluctuation
region''\cite{titov,deych0,deych}.

It might be conjectured that, similar to random lasing,
the electron transmission within the  fluctuation region is governed
by the rare events. It is easy to identify these events, which are
accidental formation of a ``minibands'' by different groups of
almost even-spaced tail states \cite{elsevier,Lifshitz}.
One can also assume that different rare events are statistically
independent. Unlike random lasing from the samples of a given area,
this assumption is by no means trivial. However, if we adopt this
assumption, then we immediately come to the conclusion that the
transmission distribution is governed by Eq. (\ref{mainQ}),
and thus can be described in a universal manner\cite{deych0}.
The analogy
between the random lasing from a finite-area sample, and transmission
of a long enough chain is based on the fact that lasing threshold
is governed by the highest-quality resonator present in the sample,
whereas the electron transmission occurs through the
``most even-spaced'' sequence of resonant tail states.

The validity of the conjecture about statistical independence of
rare events within a long disordered chain can be checked by comparison
of the  consequences of this conjecture to the numerical results.
% for the energies in the tail
%of the density of states. The relevant ``rare event'' in this problem
%is occurrence of a region within the chain with anomalously high
%transmission due to e.g.
Recently, the statistics of log-transmission
coefficients (more precisely, of the Lyapunov exponents
$\tilde{\gamma} \propto |\ln T|$) in the fluctuation tail of the
Anderson model was studied numerically in
Refs.~\onlinecite{deych0,deych}.
On the basis of the simulations performed, the authors have found
that the second, $\sigma^2=\left\langle [\tilde{\gamma} -
\langle \tilde{\gamma} \rangle ]^2 \right\rangle$,
 and the third, $\varrho =\left\langle [\tilde{\gamma} -
\langle \tilde{\gamma} \rangle ]^3 \right\rangle$,
cumulants of the distribution can be approximately described
by the following expressions
\begin{equation}
\label{second}
\tau = \sigma ^2 L l_{loc} = D \kappa ^{\alpha} + \tau_{min},
\end{equation}

\begin{equation}
\label{third}
\tau_3 = \varrho L^2 l_{loc} = D_3 \kappa^{\alpha_3} + \tau_{3,min},
\end{equation}
where $l_{loc}$ is the average localization length for an
{\em infinite} chain, $\kappa = l_{loc}\sin\left[\pi{\cal N}(E)\right]$
is the scaling parameter, and $D$, $D_3$, $\alpha$, and $\alpha_3$
are numerical constants; ${\cal N}(E)$ is the number of states per unit
length between the energy $E<0$ and the boundary $E= -2$ of the
spectrum in the absence of disorder.
Upon excluding $\kappa$ from Eqs. (\ref{second}) and (\ref{third}),
we obtain

%\begin{equation}
%\label{tau}
%\left( \frac{\tau -\tau_{min}}{D}\right)^{1/\alpha} =
% \left( \frac{\tau_3 -\tau_{3,min}}{D_3}\right)^{1/\alpha_3}
%\end{equation}

\begin{equation}
\label{tau}
\frac{\tau_3 -\tau_{3,min}}
{\left(\tau -\tau_{min}\right)^{\alpha_3/\alpha}} =
\frac{D_3}{D^{\alpha_3/\alpha}}
\end{equation}

Assume now that the {\em full} distribution of the Lyapunov exponents
is given by Eq.~(\ref{mainQ}).
This assumption suggests that the relation
Eq.~(\ref{tau}) is nothing but a different form of the relation
Eq.~(\ref{relation}). To check whether this is the case, we
express $\tau$ and $\tau_3$ in terms of parameter $\beta_s$ as
follows
\begin{equation}
\tau = \frac{\Psi^{(1)}}{\gamma \beta_s},
\end{equation}

\begin{equation}
\tau_3 = \frac{\Psi^{(2)}}{\gamma \beta^2_s}<0.
\end{equation}

Then the relation Eq.~(\ref{relation}) takes the form

\begin{equation}
\label{relation_tau}
\left| \frac{\tau_3}{\tau^{2}}  \right| = -
     \frac{\gamma \Psi^{(2)}}{\left[\Psi^{(1)}\right]^{2}} =
   \frac{72\gamma \zeta(3)}{\pi^4} \approx 0.51  .
\end{equation}
On the other hand, the simulations of Ref.~\onlinecite{deych}
yielded the following values of the parameters $\alpha =0.27$,
$\alpha_3 =0.52$, $D_3=0.73$, $D=1.27$. Then for the ratio
$\alpha_3/\alpha$ we get $\alpha_3/\alpha = 1.93$, whereas
the combination $D_3/D^{\alpha_3/\alpha}$ is approximately equal to
$0.46$. We conclude that the numerical results of
Ref. \onlinecite{deych} satisfy the analytical relation
(\ref{relation}) with high accuracy, suggesting that the distribution
Eq.~(\ref{mainQ}) might indeed describe the statistics of the coherent
transmission
coefficients through the tail of 1D Anderson chain. More conclusive
judgment could be made if the moments higher than third were
extracted from numerics of Ref. \onlinecite{deych}. In particular,
the following analytical prediction for the {\em fourth} cumulant
follows from the distribution (\ref{mainQ}) and
Eqs.~(A1), (A2) obtained using
this distribution

\begin{equation}
\label{fourth}
\frac{\tau_4}{\tau^3}=
 \frac{\gamma^2 \Psi^{(3)}}{\left[\Psi^{(1)}\right]^{3}} =
   \frac{72\gamma^2 }{5\pi^2} \approx 0.84  .
\end{equation}
Note in conclusion, that according to Eq. (\ref{mainQ}) the cumulants,
$\tau_i$ should exhibit a logarithmical dependence of the length chain,
$L$, through the parameter
$\beta_s \propto \left[\ln L \right]^{\left(\lambda -1\right)/\lambda}$.
No such dependence was revealed in the simulations\cite{deych0,deych}.
On the other hand two orders of magnitude change of $L$ in the
simulations\cite{deych} might be insufficient to uncover the above weak
dependence, especially if $\lambda$ is close to $1$.

%\acknowledgments

\vspace{8mm}
\centerline{ {\Large \bf Acknowledgments }}
\vspace{5mm}

We acknowledge the support of the National Science Foundation
under Grant No. DMR-0202790 and of the Petroleum Research
Fund under Grant No. 37890-AC6.

%\appendix

\vspace{8mm}
\centerline{ {\Large \bf Appendix A: Moments of $F_S(I)$}}
\vspace{5mm}

%average $\left\langle
%\ln^n \frac{I}{I_S} \right\rangle$.
%nowing these average, the cummulants $C_n$ can be found easily.
It is convenient to express the moments $M_n$ [Eq.~(\ref{moments})]
of the distribution function $F_S(I)$ through the average of
$\ln^k(I/I_S)$ as follows
\begin{equation}
M_n =  \left\langle \ln^n \frac{I}{\left\langle I \right\rangle }
        \right\rangle =
 \left\langle \left[ \ln \frac{I}{I_S}+\ln
\frac{I_S}{\left\langle I \right\rangle }
\right]^n \right\rangle   = \sum_{k=0}^{n} C_{k}^{n}
\left\langle \ln^k \frac{I}{I_S} \right\rangle
 \left( \ln \frac{I_S}{\left\langle I \right\rangle }  \right)^{n-k} ,
\label{average_M}
\end{equation}
where $C_k^n$ are binomial coefficients. Taking into account the
explicit expression Eq.~(\ref{main}) for $F_S(I)$, we can obtain the
average of $\ln^k(I/I_S)$ in terms of derivatives of Gamma function
\begin{eqnarray}
\left\langle \ln^n \frac{I}{I_S} \right\rangle   &  = &
  \int_0^{\infty} dI F_S(I) \ln^n \frac{I}{I_S}   =
\left( -\frac{1}{\beta_s}\right)^n
 \int_0^{\infty } dt~ e^{-t} \ln^n t    \nonumber \\
 &  = &  \left( -\frac{1}{\beta_s}\right)^n
  \left.  \left\{ \frac{d^n}{d\epsilon^n}
                         \int_0^{\infty}dt ~ t^{\epsilon }  e^{-t}
\right\}
  \right|_{\epsilon \rightarrow 0}  =
                         \left( -\frac{1}{\beta_s}\right)^n
  \left.  \left\{
    \frac{d^n}{d\epsilon^n} \Gamma(1+\epsilon) \right\}
  \right|_{\epsilon \rightarrow 0}  ,
\label{average_S}
\end{eqnarray}
where the variable $t = (I/I_S)^{-\beta_s}$ has been introduced in the
intermediate integrals. It is convenient to rewrite the equation
(\ref{average_S}) through the digamma function $\Psi(x) = d\ln \Gamma(x)/dx$
as follows
\begin{eqnarray}
\left\langle \ln^n \frac{I}{I_S} \right\rangle   &  = &
                                  \left( -\frac{1}{\beta_s}\right)^n
  \left. \left\{ \frac{d^n}{d\epsilon^n}
                                  \exp \left[ \sum_{k=0}^{\infty}
  \frac{\Psi^{(k)}}{k!} \int_0^{\epsilon } dv~ v^k
   \right] \right\}
   \right|_{\epsilon \rightarrow 0}     \nonumber \\
   & &  =   \left( -\frac{1}{\beta_s}\right)^n
  \left.  \left\{ \frac{d^n}{d\epsilon^n}
                                  \exp \left[ \sum_{k=0}^{\infty}
  \frac{\Psi^{(k)}}{(k+1)!} \epsilon^{k+1}
   \right] \right\}
   \right|_{\epsilon \rightarrow 0} .
\label{average_S1}
\end{eqnarray}
Here the abbreviations $\Psi^{(k)}  = \left. d^k\Psi(x)/dx^k \right|_{x=1} $
have been introduced. Substituting Eq.~(\ref{average_S1}) into
Eq.~(\ref{average_M}), we obtain the expressions for the first
three moments
\begin{equation}
M_1 = -\frac{\Psi^{(0)}}{\beta_s} - \ln \left[ \Gamma
     \left(1-\beta_s^{-1} \right)\right] ,
\end{equation}

\begin{equation}
M_2 = \frac{\Psi^{(1)}}{\beta^2_s} + M_1^2  ,
\end{equation}
and
\begin{equation}
M_3 = -\frac{\Psi^{(2)}}{\beta^3_s} + 3 M_1 M_2 -2 M_1^3    ,
\end{equation}
where $\Psi^{(0)} = \Psi(1) = -\gamma$, $\Psi^{(1)} = \pi^2/6$, and
$\Psi^{(2)} = -2\zeta (3)$. Here $\zeta (x)$ is the zeta-function.

%\end{document}

\begin{figure}
%\narrowtext
\centerline{
\epsfxsize=4.5in
\epsfbox{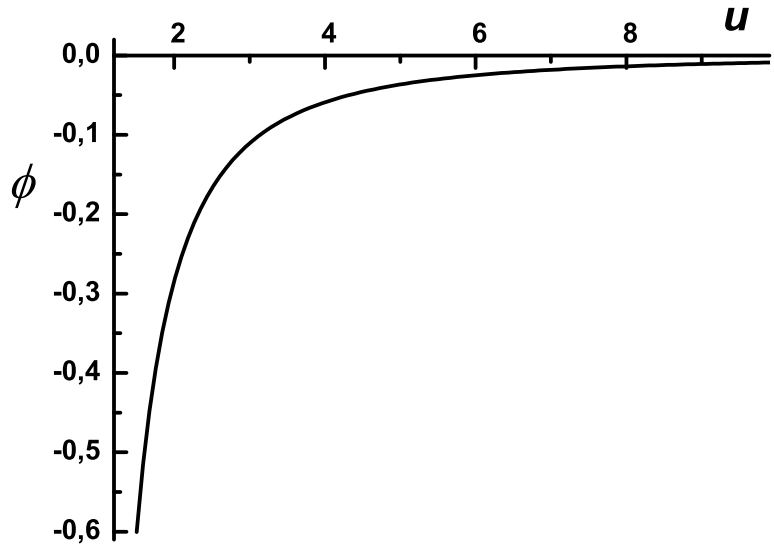}
\protect\vspace*{0.1in}}
\protect\caption[sample]
{\sloppy{ \large Dimensionless function $\phi(u)$ is plotted from Eq.
(\ref{phi}).}}
\label{figone}
\end{figure}

\begin{figure}
%\narrowtext
\centerline{
\epsfxsize=4.5in
\epsfbox{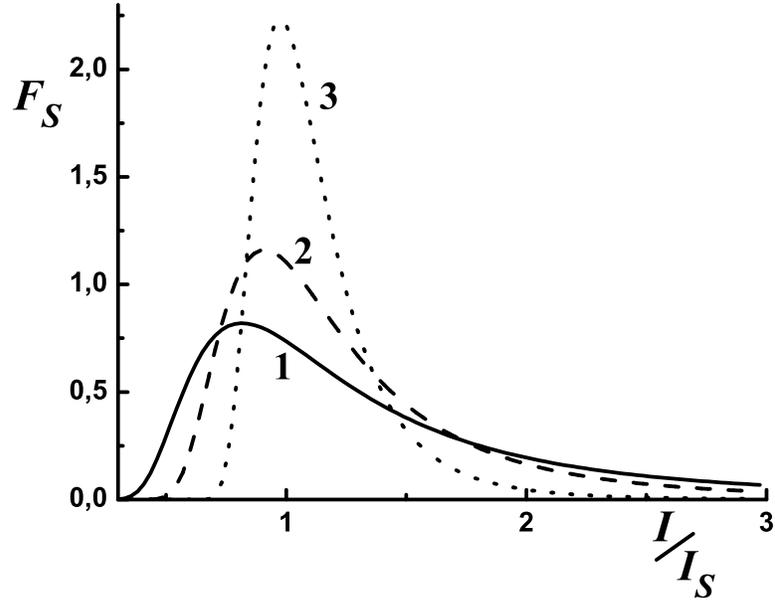}
\protect\vspace*{0.1in}}
\protect\caption[sample]
{\sloppy{ \large Distribution function of the lasing thresholds over
samples (excitation spots) calculated from Eq. (\ref{main}) is
plotted for different values of parameter $\beta_s$. Solid line:
$\beta_s=2$; dashed line: $\beta_s=4$; dotted line: $\beta_s=6$.
}}
\label{figtwo}
\end{figure}

\end{document}